\documentclass[pre,twocolumn]{revtex4-2}
\usepackage{dcolumn}
\usepackage{graphicx}
\usepackage{epsfig}
\usepackage{amsmath}
\usepackage{amsthm}
\usepackage{amsfonts}
\usepackage{amssymb}
\usepackage{bm}
\usepackage{times}
\usepackage{xcolor}

\newtheorem{lem}{Lemma}

\theoremstyle{definition}

\newcommand{\Tr}{\mathrm{Tr}}
\newcommand{\QKL}{S(\gamma_1\rVert\gamma_0)}
\newcommand{\KL}{D_\text{KL}}
\newcommand{\E}{\mathbb{E}}
\newcommand{\OP}{\mathcal{O}}
\newcommand{\M}{\mathcal{M}}
\newcommand{\Ham}{\mathcal{H}}

\newcommand{\bra}[1]{\langle #1 |}
\newcommand{\ket}[1]{| #1 \rangle}
\begin{document}
\title{Quantum Fluctuation-Response Inequality and Its Application in Quantum Hypothesis Testing}
\author{Yan Wang}
\email{wangyan@wayne.edu}
\affiliation{Department of Mathematics, Wayne  State University, Detroit, Michigan 48202, USA}
\affiliation{Institute for AI and Data Science (AIDaS), Wayne  State University, Detroit, Michigan 48202, USA}

\begin{abstract}
We establish a general quantum fluctuation-response inequality that bounds the difference in expectation values of an observable between two quantum states via the quantum relative entropy. For observables with bounded spectra, we further strengthen this result by exploiting the sub-Gaussian property, and explicitly relate our bound to the sub-Gaussian norm of the observable. This allows us to derive a novel non-asymptotic bound on the sum of statistical errors in quantum hypothesis testing, which complements and can be more informative than existing bounds. We also demonstrate the versatility of our results through problems such as thermodynamic hypothesis testing and quantum speed limits in physics, and generalizability of algorithms in machine learning.
\end{abstract}
\maketitle
\section{Introduction}\label{sec:Intro}
Quantifying the mean difference of a quantity at two states is of great interest in both classical and quantum settings, and in various fields from physics to machine learning and statistics. For example, in physics, when some system parameter is perturbed from $\theta_0$ to $\theta=\theta_0+\delta\theta$, the way the system responds to such a change reveals important properties of the system. When $\delta\theta$ is small, the quantitative difference between $\Tr (\OP\gamma_{\theta_0})$ and $\Tr (\OP\gamma_{\theta})$ is captured by the celebrated linear response theory \cite{Book:Kubo}, where $\OP$ is some observable of interest and $\gamma_{\theta_0}$ and $\gamma_\theta$ are the corresponding density operators of the original and the perturbed system, respectively. In machine learning, one central task is to upper bound the difference between the empirical risk $\mathbb{E}_{P_n}L(h,Z)$ and the population risk $\mathbb{E}_{P_Z}L(h,Z)$ incurred by a regression function $h$ on test data $Z$ \cite{AoS:Bartlett}, where $L$ is a given loss function, $P_Z$ is the underlying data-generating distribution, and $P_n$ is the empirical distribution of training data. A learning algorithm resulting in regression functions with a small difference between the population and empirical risks is highly favored. In statistics, for a test function $\Phi$, $\E_{H_0}\Phi=\alpha$ is the type I error (false positive) rate, and  $1-\E_{H_1}\Phi=\beta$ gives the type II error (false negative) rate, where $\E_{H_0}$ and $\E_{H_1}$ denote the expectations with respect to the null hypothesis $H_0$ and the alternative hypothesis $H_1$, respectively \cite{Book:Keener}. Note that $\E_{H_0}\Phi - \E_{H_1}\Phi$ is equal to the sum of error rates $\alpha+\beta$ under the test function $\Phi$, up to a constant 1. Thus, quantifying this difference is crucial in seeking a good test function with small statistical error rates \cite{Me:ISIT}. At a high level, all such scenarios boil down to the understanding of key factors that determine the difference:
\begin{align*}
|\E_0 \mathcal{O} - \E_1\mathcal{O}| \lesssim f(\text{system, operation}).
\end{align*}
In general, the mean difference (classical or quantum) of some quantity $\mathcal{O}$ at two different states is controlled by both the system properties and the way we manipulate the system. In linear response, such a mean difference is equal to $\chi_{\OP}\delta\theta$ with $\delta \theta$ quantifying to what extent the system is modified experimentally and the associated susceptibility $\chi_{\OP}$ reflecting the thermodynamic property of the original system \cite{Book:Kubo}. In machine learning and statistics, the results typically take the form of an \emph{inequality}, in terms of some distance measure between systems, which are characterized by two probabilities ($P_Z$ and $P_n$ in machine learning, and $P_0$ and $P_1$ in statistics which represent data distributions under $H_0$ and $H_1$, respectively) and some complexity measure of a set of possible operations, usually represented by a given function class \cite{AoS:Bartlett,Me:ISIT,IEEE:Zou,NIPS:Raginsky}. 

In physics, a similar inequality was developed recently, and an elegant framework has been established for analyzing the mean difference of a generic quantity between an arbitrary perturbed state (represented by $P_1$) and a reference state (represented by $P_0$) in the classical setting, known as the fluctuation-response inequality (FRI) \cite{PNAS:Sasa}. Since $P_0$ and $P_1$ can describe rather general nonequilibrium states, the broad applicability of the FRI is highly appealing. There are two key ingredients of the theory. The first is the relative entropy (or the Kullback-Leibler divergence) $\KL(P_1\Vert P_0)$, which quantifies the difference between the perturbed and reference systems. The second is the moment-generating function (MGF) $\E_{P_0}e^{s(\OP-\E_{P_0}\OP)}$, which encodes the distributional information of the reference system that can be accessed by the operation of measuring a quantity $\OP$. When $\OP$ follows a Gaussian distribution in the reference system, its variance fully determines the MGF, and the FRI connects the mean difference directly to the variance (fluctuation around the mean with respect to $P_0$) and the relative entropy. If the distribution of $\OP$ under $P_0$ belongs to a broader class beyond Gaussian, such as the sub-Gaussian family, then another clean result can be obtained by replacing the variance by a sub-Gaussian parameter \cite{Me:PRE}. In both cases, the FRI provides a principled upper bound on the mean difference:
\[
|\E_{P_0} \OP - \E_{P_1}\OP|\leq \sqrt{2\sigma_0^2\KL(P_1\Vert P_0)},
\]
where $\sigma_0^2$ denotes the sub-Gaussian parameter of $\OP$ under $P_0$ ($\sigma_0^2$ becomes the variance when $P_0$ is Gaussian).

One might naturally want to extend such a result to the quantum regime. However, the methodology underlying the development of the classical FRI is no longer applicable in the quantum case because density operators, which are the quantum counterparts of classical distributions, may be noncommutative. To overcome this limitation, we develop new techniques in this work, based on which we nontrivially establish a similar framework in the quantum regime. We first establish a quantum version of the FRI, which bounds the mean difference of a generic observable at two arbitrary quantum states in terms of their quantum relative entropy. Second, by taking advantage of the sub-Gaussian property associated with any measurement plan, we provide an operationally useful bound of error rates for quantum hypothesis testing \cite{Book:Hayashi}, which is a non-asymptotic result that can be applied to any finite number of copies of the system. The bound complements existing bounds and we show that it can be more informative in certain cases. It is worth emphasizing that, while the technique used in this work is closely related in spirit to the Gibbs variational principle and variational representations of quantum relative entropy, the primary contributions of this work lie in the development of a quantum FRI, especially its sub-Gaussian formulation, and its application to quantum hypothesis testing. We also discuss possible applications of our results in problems related to thermodynamic inference \cite{ARCMP:Seifert} and the speed limit \cite{PRX:SL} in physics, as well as the generalizability of algorithms in machine learning. We believe that, given the similarity between the goals of statistical physics and other disciplines as discussed above, our results are expected to have broad applicability across various scientific fields. 

In the remainder of this work, we establish the quantum fluctuation-response inequality and its sub-Gaussian version in Sec.~\ref{sec:QFRI}. We then apply these results to quantum hypothesis testing in Sec.~\ref{sec:QHT}, and discuss further applications in Sec.~\ref{sec:more_appl}. Finally, we conclude in Sec.~\ref{sec:conclusion}.

\section{Quantum fluctuation-response inequality} \label{sec:QFRI}
Let $\gamma_0$ and $\gamma_1$ be two density operators that describe a system at two different conditions, where $\gamma_0$ represents the reference system, and $\gamma_1$ the ``perturbed'' system. Our main results will be expressed in terms of the quantum relative entropy $S(\gamma_1\Vert\gamma_0)$ \cite{Book:Hayashi,Book:Mike_n_Ike} , which is defined by 
\begin{align*}
    S(\gamma_1\Vert\gamma_0) \equiv \Tr(\gamma_1\ln\gamma_1) - \Tr(\gamma_1\ln\gamma_0)
\end{align*}
if the support of $\gamma_1$ is contained in the support of $\gamma_0$. Otherwise, $S(\gamma_1\Vert\gamma_0)=+\infty$. It is known that $S(\gamma_1\Vert\gamma_0)\geq 0$ and the equality holds only when $\gamma_1=\gamma_0$. Hence $S(\gamma_1\Vert\gamma_0)$ naturally serves as a measure to quantify the difference from the perturbed system $\gamma_1$ to the reference system $\gamma_0$. We will proceed with the nontrivial case that $S(\gamma_1\Vert\gamma_0)<+\infty$. We will also evaluate an operator function such as $\ln(\cdot)$ only on the support, where the operator in question has strictly positive eigenvalues. For a generic Hermitian operator $\mathcal{O}$ of interest, we want to bound $|\Tr(\mathcal{O}\gamma_0)-\Tr(\mathcal{O}\gamma_1)|$ and generalize the classical FRI \cite{PNAS:Sasa}. 

\subsection{Derivation of quantum FRI}
The main challenge in extending the classical fluctuation-response theory to the quantum regime stems from the potential noncommutativity of density operators. That is, in general, $\gamma_0$ and $\gamma_1$ may not be simultaneously diagonalized using the same set of eigenvectors. We tackle this problem by carefully constructing useful ancillary operators. 

First, let the ``centered'' version of $\OP$ (with respect to $\gamma_0$) be
\[
\OP_c\equiv \OP-\Tr(\OP\gamma_0)I,
\]
where $I$ is the identity operator in the Hilbert space of interest. Note that $\Tr(\gamma_0)=1$ (since $\gamma_0$ is a density operator), we immediately have that $\Tr(\OP_c\gamma_0)=0$. 

Next, we introduce an ancillary positive parameter $s$ to derive a tight bound for $|\Tr(\gamma_1\OP) - \Tr(\gamma_0 \OP)|$. This is a standard approach in probability \cite{Book:Roman}. For any $s>0$, we define:
\begin{align}
& B \equiv \xi s \OP_c + \ln\gamma_0, \label{eq:B}\\
& A \equiv \ln\gamma_1 + \lambda I, \label{eq:A}
\end{align}
where 
\begin{align*}
\lambda &=\Tr \big(\gamma_1(B-\ln\gamma_1)\big), \text{ and}\\
\xi&=\text{sgn}\big(\Tr(\OP\gamma_1) - \Tr(\OP\gamma_0) \big)
\end{align*}
with $\text{sgn}(\cdot)$ being the sign function. One can verify that $B$ and $A$ are Hermitian and 
\begin{align}
\Tr\big((B-A)\gamma_1\big)=0. \label{eq:TrB_A1} 
\end{align}
To proceed, let us recall Klein's inequality, which is a property of convex functions in the quantum regime, and by which we have 
\[
\Tr(e^B) \geq \Tr(e^A)+\Tr(e^A(B-A)).
\]
Inserting Eq. (\ref{eq:A}) into the expression, and noting Eq. (\ref{eq:TrB_A1}), the definition of $\lambda$, and the fact that the commutator $[\ln\gamma_1,I]=0$, we have 
\begin{align}
\Tr(e^B) &\geq \Tr\left(e^{\ln\gamma_1+\lambda I}\right) + \Tr\left(e^{\ln\gamma_1+\lambda I}(B-A)\right) \notag\\
            &=  \exp\left(\Tr \big(\gamma_1(B-\ln\gamma_1)\big)\right).\notag
\end{align}
Further inserting Eq. (\ref{eq:B}) to the expression above and taking the logarithm on both sides yields 
\begin{align}
\ln\Tr(e^{\xi s\OP_c+\ln\gamma_0}) 
&\geq \xi s[\Tr(\gamma_1\OP)-\Tr(\gamma_0\OP)] - \QKL\notag\\
&= s|\Tr(\gamma_1\OP)-\Tr(\gamma_0\OP)| - \QKL.\notag
\end{align}  
Details of the derivation under a more general framework are provided in Appendix \ref{Appendix:Derivation_QFRI}. Here, we have tacitly assumed that in the inequality each involved operator is trace-class. This assumption is mild and holds in most practical settings. For many operators of interest, such as those on a finite-dimensional Hilbert space, the above inequality holds for all $s>0$ \cite{Note}. Hence, we might take the infimum with respect to $s$ to obtain
\begin{align}\label{eq:QFR}
|\Tr(\gamma_1\OP)- & \Tr(\gamma_0\OP)| \notag\\
                              &  \leq \inf_{s>0}\frac{1}{s}\left[\ln\Tr(e^{\xi s\OP_c+\ln\gamma_0})+\QKL\right]\notag\\
                              & \leq \inf_{s>0}\frac{1}{s}\left[\ln\Tr(\gamma_0 e^{\xi s\OP_c})+\QKL\right],
\end{align} 
where the second inequality is due to the Golden-Thompson inequality \cite{Thompson,Golden}. This is the first main result of this work, which may be termed as the quantum fluctuation-response inequality (QFRI). 

Following similar steps, we can also obtain a one-sided version of the above result, based on which we recover the Gibbs variational principle (at temperature $T$ and the Boltzmann constant is set to $k_B=1$)
\[
-\ln\Tr(e^{-\Ham/T}) \le \Tr(\gamma\Ham/T) + \Tr(\gamma\ln\gamma),
\] 
where $\Ham$ is the Hamiltonian of a physical system and $\gamma$ is an arbitrary density operator. Equality is attained when $\gamma$ represents the thermal state, i.e., when $\gamma = e^{-\Ham/T}/\Tr(e^{-\Ham/T})$. (Technical details are given in Appendix \ref{Appendix:Derivation_QFRI}.) This byproduct further reveals the relevance of our work to statistical physics.

Formally, the QFRI (\ref{eq:QFR}) is similar to its classical counterpart \cite{PNAS:Sasa}. By replacing density operators by probabilities, the quantum relative entropy by the classical relative entropy, and the quantum operator by a random variable, the QFRI reduces to the classical FRI. However, despite the formal similarity, there is an important difference between them. In the classical case, the events that probabilities are assigned to can be trajectories of the forward and backward processes as in the context of stochastic thermodynamics \cite{RPP:Seifert}. While, in the quantum case, a density operator is used to describe the state of a physical system, rather than its evolution. Hence, in this sense, our result (\ref{eq:QFR}) does not completely correspond to the classical FRI. However, we expect that as long as quantum states are concerned, (\ref{eq:QFR}) is a natural generalization of its classical counterpart and is of great relevance in statistical physics, especially in the nonlinear response regime. In linear response, only the second moment of $\OP_c$ is involved. (See discussions below.) Our result involves all moments of $\OP_c$ (or the full counting statistics \cite{RMP:FCS}) that contain much richer ``distributional'' information of the quantum state $\gamma_0$ relevant to the observable $\OP$. By expanding $\ln\Tr(\gamma_0 e^{\xi s \OP_c})$ into a series and retaining higher-order terms to the desired accuracy, one can establish a hierarchy of bounds in the nonlinear response regime perturbatively.  

However, for a wide class of operators, in particular those defined in a finite-dimensional Hilbert space, we can go beyond (\ref{eq:QFR}) by taking advantage of the sub-Gaussian property. In this case, we use a \emph{single} quantity $\sigma_{OP_0}$ defined below to capture the essential information of $\gamma_0$ relevant to the observable $\OP$, which is alternative to the perturbation approach using the full counting statistics.

\subsection{Sub-Gaussian QFRI}\label{subsec:subGQFRI}
We begin by introducing classical sub-Gaussian random variables, which are widely used in statistics and machine learning. A sub-Gaussian random variable generalizes the Gaussian in the sense that its tail probability decays at a rate similar to that of a Gaussian \cite{Book:Martin,Book:Roman}. Precisely, a random variable $X$ that follows distribution $P$ is said to be sub-Gaussian if there exists some parameter $\sigma^2$ such that
\begin{align}\label{eq:subG}
\E_{X\sim P} \exp[t(X-\E_{X\sim P}X)] \leq e^{\sigma^2t^2/2}
\end{align}  
for all $t\in\mathbb{R}$. The square root of the infimum of all such $\sigma^2$ is defined to be the sub-Gaussian norm of $X$, denoted $\sigma_{XP}$ \cite{subGnorm,Me:PRE}. Sometimes, $\sigma_{XP}^2$ is also referred to as the variance proxy and it is always no less than the variance of $X$ under $P$. Many common distributions are sub-Gaussian, including the Gaussian itself. In particular, all \emph{bounded} random variables are sub-Gaussian. This fact is particularly relevant in quantum information science where a finite dimensional Hilbert space is usually considered, and physical quantities of interest are typically bounded \cite{Book:Hayashi,Book:Mike_n_Ike}. The sub-Gaussian concept is extremely helpful in such cases. 

It is known that if $X$ is bounded in $[a,b]$, then $\sigma_{XP}\leq(b-a)/2$ holds universally, irrespective of the details of the distribution $P$ \cite{Book:Martin,Book:Roman}. However, given $P$, one can numerically calculate $\sigma_{XP}$ in a principled way or find a more informative upper bound than $(b-a)/2$. This will be discussed below in the context of quantum hypothesis testing.

To further take advantage of the sub-Gaussianity in (\ref{eq:QFR}), we use an orthonormal basis of the Hilbert space in question that diagonalizes $\OP_c$, such that 
\[
\Tr(\gamma_0e^{\xi s \OP_c})=\sum_{i}\langle i| \gamma_0 |i\rangle e^{\xi s o_i},
\]
where the eigendecomposition of $\OP_c$ is given by $\OP_c=\sum_{i}o_i|i\rangle\langle i|$. Since $\gamma_0$ is a density operator, it is legitimate to interpret  
\[
P_0=\{p_i\equiv\langle i| \gamma_0 |i\rangle\}
\]
as a probability distribution with $p_i\in[0,1]$ for all $i$. We may also define $O_c$ as the corresponding random variable that follows the distribution $P_0$, which equals $o_i$ with probability $p_i$. This way, we rewrite the trace as an expectation: 
\[
\Tr(\gamma_0e^{\xi s \OP_c}) = \E_{O_c\sim P_0} \exp(\xi s O_c).
\]
Note that $\E_{O_c\sim P_0}O_c=0$. Hence if the random variable $O_c$ is sub-Gaussian with sub-Gaussian norm $\sigma_{OP_0}$, then by (\ref{eq:subG}), we obtain
\begin{align}
 \Tr(\gamma_0e^{\xi s \OP_c}) \leq \exp[\sigma_{OP_0}^2(\xi s)^2/2] = \exp(\sigma_{OP_0}^2s^2/2).\notag
\end{align}
Inserting this to (\ref{eq:QFR}) yields 
\begin{align}\label{eq:subGQFR}
|\Tr(\gamma_1\OP)-\Tr(\gamma_0\OP)| 
&\leq \inf_{s>0}\frac{1}{s}\left[\ln e^{\sigma_{OP_0}^2s^2/2}+\QKL\right]\notag\\
&= \sigma_{OP_0}\sqrt{2\QKL},
\end{align}
where the infimum is achieved at $s=\sqrt{2\QKL}/\sigma_{OP_0}$. This sub-Gaussian QFRI is our second main result, which is more relevant than (\ref{eq:QFR}) to cases where the deviation from the linear response regime is mild and the operator of interest has a bounded spectrum.

In the linear regime, one can verify that (\ref{eq:subGQFR}) covers the result from the linear response theory. Suppose $\Ham_0$ is the Hamiltonian of the reference system and $\Ham_1=\Ham_0+\varepsilon \OP$ is the Hamiltonian of the perturbed system, where $|\varepsilon|\ll 1$ is a smallness parameter. Then $\gamma_{i}=e^{-\beta_0 \Ham_i}/Z_i$ for $i=0,1$, where $\beta_0$ is the inverse temperature at both states  and $Z_i=\Tr(e^{-\beta_0 \Ham_i})$ is the corresponding partition function. A standard result from the linear response theory gives
\[
|\Tr(\OP\gamma_1)-\Tr(\OP\gamma_0)|\approx |\epsilon| \beta_0 \mathrm{Var}_{\gamma_0}[\OP],
\]
where $\mathrm{Var}_{\gamma_0}[\OP]=\Tr(\gamma_0 \OP^2)-(\Tr(\gamma_0\OP))^2$ is the variance of $\OP$ under $\gamma_0$. On the other hand, it is also straightforward to find $S(\gamma_1\rVert\gamma_0)\approx \frac{1}{2}\varepsilon^2 \beta_0^2 \mathrm{Var}_{\gamma_0}[\OP]$ and $\ln\Tr(\gamma_0e^{\xi s\OP_c}) \approx \frac{1}{2}s^2 \mathrm{Var}_{\gamma_0}[\OP]$, assuming $s$ is small. Thus, based on our theory, we have
\begin{align*}
|\Tr(\OP\gamma_1)& -\Tr(\OP\gamma_0)|\\
&\lesssim \inf_{s>0}\frac{1}{s}\left[\frac{1}{2}s^2 \mathrm{Var}_{\gamma_0}[\OP] + \frac{1}{2}\varepsilon^2 \beta_0^2 \mathrm{Var}_{\gamma_0}[\OP]\right]\\
& = |\varepsilon| \beta_0 \mathrm{Var}_{\gamma_0}[\OP].
\end{align*}
The infimum is achieved when $s=|\varepsilon|\beta_0$, which is small for weak perturbation $|\varepsilon|\ll 1$ and high temperature $\beta_0\ll 1$, thus justifying the above assumption that $s$ is small. This result is consistent with the linear response theory. In fact, in this case, the variance proxy $\sigma^2_{OP_0}$ is approximated by the variance $\mathrm{Var}_{\gamma_0}[\OP]$ and the quantum relative entropy is also captured by the variance. Hence (\ref{eq:subGQFR}) naturally serves as a generalization of the linear response result in the sub-Gaussian nonlinear regime. This complements the perturbative approach to nonlinear response by including more terms in the expansion of $\ln\Tr(\gamma_0 e^{\xi s \OP_c})$.

Moreover, new bounds of possible physical interest can be obtained by (\ref{eq:subGQFR}). For example, let us pick $\OP=-\Ham$ with $\Ham$ being the Hamiltonian. Without loss of generality, the eigenvalues of $\Ham$ are supposed to be bounded in $[0,H_{\max}]$ for a finite system. Then (\ref{eq:subGQFR}) implies that $|U_1-U_0|\leq \sigma_{HP_0}\sqrt{2\QKL}$ with $\sigma_{HP_0}\leq H_{\max}/2$ being the sub-Gaussian norm of $\Ham$ with respect to $\gamma_0$ and $U_{0,1}=\Tr(\gamma_{0,1} H)$ being the corresponding ``internal energies'' of the system at $\rho_{0,1}$. Hence the mean energy difference at different states can be bounded in terms of the quantum relative entropy between the states. If we further let $\gamma_0=e^{-\Ham/T}/\Tr(e^{-\Ham/T})$ be the thermal state at temperature $T$, and define $F_{0,1}\equiv U_{0,1}-TS_{0,1}$ as ``Helmholtz free energies'' at states $\rho_{0,1}$, respectively, with $S_{0,1}=-\Tr(\gamma_{0,1}\ln\gamma_{0,1})$ being entropies, then by these definitions we have 
\begin{align*}
	&(F_1-F_0)/T \notag\\
	={}& \Tr\left((\gamma_1-\gamma_0)\Ham/T\right)+\Tr(\gamma_1\ln\gamma_1) - \Tr(\gamma_0\ln\gamma_0)\\
	={}&\Tr((\gamma_1-\gamma_0)\Ham/T) +\Tr((\gamma_1-\gamma_0)\ln\gamma_0)+\QKL.
\end{align*}
Since $\gamma_0=e^{-\Ham/T}/\Tr(e^{-\Ham/T})$, it is straightforward to find that
\(
\Tr((\gamma_1-\gamma_0)\ln\gamma_0) = -\Tr((\gamma_1-\gamma_0)\Ham/T)-\ln\Tr(e^{-\Ham/T})\Tr(\gamma_1-\gamma_0) = -\Tr((\gamma_1-\gamma_0)\Ham/T)
\),
which thus yields $(F_1-F_0)/T = \QKL$, and we finally have
\[
|U_1-U_0|\leq \sigma_{HP_0}\sqrt{2(F_1-F_0)/T}.
\]
Note that $T$ is fixed in this case, and $F_0$ is minimized at equilibrium. Hence $F_1$ is greater than $F_0$ in general, and the above bound is physically valid. 

We believe that, as general upper bounds, the QFRI (\ref{eq:QFR}) and its sub-Gaussian version (\ref{eq:subGQFR}) can yield physically insightful results when applied to specific problems. But beyond statistical physics, they are also useful in other fields such as statistics and machine learning. In the following, we focus on the application of (\ref{eq:subGQFR}) in quantum hypothesis testing where $\OP_c$ is constructed from a measurement plan.

\section{Quantum hypothesis testing}\label{sec:QHT}
Classical statistics deals with the discrimination of two probabilities, while quantum hypothesis testing aims to discriminate two quantum states $\gamma_0$ and $\gamma_1$ with the null and alternative hypotheses being:
\[
H_0:\gamma = \gamma_0 \text{ vs. } H_1: \gamma=\gamma_1. 
\]
From the classical to the quantum setting, some similar ideas are shared in constructing optimal tests, and some classical results have their quantum counterparts with almost the same form \cite{PRL:Chernoff,AoS:Chernoff,TIT:Stein}. However, one fundamental difference is that in the latter case, in addition to the quantum state, the measurement plan also plays a role in determining the outcome distribution. This additional factor in quantum hypothesis testing has no classical counterpart. Also, in the quantum case, one can incorporate the test step into the measurement step to compactly design a positive operator-valued measurement (POVM) $\mathbb{M}=\{\M_0,\M_1\}$, where $\M_{0}$ and $\M_1$ are positive semidefinite operators with $\M_0+\M_1=I$. The operator $\M_0$ corresponds to the outcome equal to 0, meaning $\gamma_0$ is accepted; similarly, $\M_1$ corresponds to the outcome 1 which means $\gamma_1$ is accepted. Given $n$ copies of the system, whose state is either described by the tensor product of $\gamma_0$ as $\gamma_0^{\otimes n}$, or by that of $\gamma_1$ as $\gamma_1^{\otimes n}$, the error rates are calculated as $\alpha=\Tr(\gamma_0^{\otimes n}\M_1)$ and $\beta=\Tr(\gamma_1^{\otimes n}\M_0)$. Similar to the classical case, it is well known that 
\begin{align}\label{eq:err_bound_opt}
\alpha+\beta\geq1 - \frac{1}{2}\lVert \gamma_0^{\otimes n}-\gamma_1^{\otimes n} \rVert_1,
\end{align}
where $\lVert\cdot\rVert_1$ denotes the trace norm, i.e., the sum of the singular values. In theory, the equality can be attained by the optimal measurement plan for the $n$-copy system $\mathbb{M}_\text{opt}$. However, the importance of this result is largely theoretical. This is because the optimal error bound and the associated measurement plan require the precise knowledge of the spectra of $\gamma_0^{\otimes n}-\gamma_1^{\otimes n}$. The corresponding computation can be prohibitive, as the cost generally grows linearly with the support size of $\gamma_0^{\otimes n}$ and thus increases exponentially fast in $n$. Alternatively, thanks to quantum Pinsker's inequality \cite{Pinsker}, another lower bound for the error rates can be found as
\begin{align}\label{eq:QPinsker}
\alpha+\beta \geq 1-\frac{1}{2}\sqrt{2nS(\gamma_1 \rVert \gamma_0)}.
\end{align}
In this case, the evaluation of the bound is much easier as we only need to calculate $S(\gamma_1\Vert\gamma_0)$ for a single copy of the system. These results provide universal lower bounds  for \emph{any} POVM used to perform the hypothesis testing. 

However, if we do not consider a generic POVM, but rather focus on some specific type of POVMs, then the universal bounds (\ref{eq:err_bound_opt}) and (\ref{eq:QPinsker}) can be further improved. We show that applying the sub-Gaussian QFRI~(\ref{eq:subGQFR}) yields a stronger and more informative bound by explicitly incorporating information about the measurement plan. This is of practical importance, as practitioners typically prefer error bounds associated with a \emph{given} measurement plan~$\mathbb{M}$.

\subsection{Measurement-dependent sub-Gaussian error bound}

Suppose $\mathbb{M}$ is a measurement plan designed for $n$ copies of the system. To establish an error bound associated with $\mathbb{M}$, we set $\OP \leftarrow \M_1$ and $\gamma_{0,1}\leftarrow\gamma_{0,1}^{\otimes{n}}$ in the sub-Gaussian QFRI (\ref{eq:subGQFR}). Since $\Tr(\gamma_0^{\otimes n}\M_1)=\alpha$ and $\Tr(\gamma_1^{\otimes n}\M_1)=\Tr(\gamma_1^{\otimes n}(I-\M_0))=1-\beta$, and by the tensorization property of the quantum relative entropy that $S(\gamma_1^{\otimes n}\Vert \gamma_0^{\otimes n})=nS(\gamma_1\Vert \gamma_0)$, we have
\begin{align}\label{eq:errsum}
\alpha + \beta \geq  1 - \sigma_{0}\sqrt{2nS(\gamma_1 \rVert \gamma_0)},
\end{align}
where $\sigma_0$ is the sub-Gaussian norm of some measurement-dependent random variable under the null hypothesis $H_0$, which will be detailed below. Abusing the notation a bit, we might also define $\sigma_0$ to be the sub-Gaussian norm of $\M_1$ with respect to the null hypothesis $H_0$. For all practically nontrivial measurement plans with $\alpha<0.5$ (a random guess results in $\alpha=0.5$), we will prove that
\begin{align}\label{eq:sigma0}
\sigma_0\leq \sqrt{\frac{\alpha-0.5}{\ln(\alpha/(1-\alpha))}}<\frac{1}{2}.
\end{align}
The error bound (\ref{eq:errsum}), together with (\ref{eq:sigma0}), constitutes the third main result of this work. This bound for $\sigma_0$ also shows that our result (\ref{eq:errsum}) is always stronger than the quantum Pinsker bound (\ref{eq:QPinsker}). In particular, the improvement is huge when $\alpha\approx 0$. This small-$\alpha$ regime corresponds to controlling the Type I error rate at a very low level, which is standard practice in hypothesis testing. Physically, this reflects the situation in which falsely rejecting $H_0$ is regarded as much more serious than failing to reject $H_0$ when $H_1$ is true.

Before we show how to calculate or bound $\sigma_0$, several points that we want to emphasize are in order. First, our bound (\ref{eq:errsum}) evidently consists of two parts. The quantum relative entropy $S(\gamma_1 \rVert \gamma_0)$ describes the distance between two quantum states, while $\sigma_0$ characterizes the performance of the POVM we use to implement the test if the system is in state $\gamma_0$. If the two states $\gamma_0$ and $\gamma_1$ are close to each other with a small $\QKL$, then intuitively it will be difficult to distinguish them and the statistical error rates will be high. On the other hand, if we want the test to be successful with high probability when the system is in state $\gamma_0$, then the resulting $\mathbb{M}$ will be designed to have a small type I error rate $\alpha$, which consequently leads to a small $\sigma_0$ by (\ref{eq:sigma0}). By our error bound (\ref{eq:errsum}), a small $\sigma_0$ increases the lower bound of the sum of error rates. Hence (\ref{eq:errsum}) quantitatively shows that by suppressing one type of error will inevitably increase the other. Based on (\ref{eq:errsum}) and (\ref{eq:sigma0}), once we know an upper bound of $\alpha=\Tr(\gamma_0^{\otimes n}M_1)$ for a given $\mathbb{M}$, a sensible lower bound can be obtained for $\beta$. The explicit dependence on $\mathbb{M}$ is a distinctive feature of our result.

Second, since $\sigma_0$ contains the information of both the system $\gamma_0$ and the measurement $\mathbb{M}$, it is expected that our result (\ref{eq:errsum}) could be more informative in some cases. For example, in the case that the type I error rate $\alpha$ is controlled below a given level, $\sigma_0$ will also be small by (\ref{eq:sigma0}). In particular, both $\alpha$ and $\sigma_0$ may converge to 0 as $n$ is increased. If one requires a specific convergence rate for $\sigma_0$, say, $\sigma_0\sim n^{-1/2}$, then (\ref{eq:errsum}) gives an almost constant error bound while other error bounds will approach 0 as $n$ is increased. A greater lower bound by our result is thus more informative than (\ref{eq:err_bound_opt}) and (\ref{eq:QPinsker}). 

Third, this result is \emph{non-asymptotic} and holds for any finite $n$. This is different than asymptotic results like the quantum Chernoff bound and quantum Stein's lemma \cite{PRL:Chernoff,AoS:Chernoff,TIT:Stein}, which only hold as $n\rightarrow\infty$. As discussed above, a finite $n$ may be more relevant when it comes to realistic settings. Our result is therefore more informative than the existing asymptotic bounds, especially when a finite-$n$ guarantee on the error rate is critical.

Fourth, by swapping $H_0$ and $H_1$, there is a natural twin of (\ref{eq:errsum}) that $ \beta + \alpha \geq  1 - \sigma_{1}\sqrt{2nS(\gamma_0 \rVert \gamma_1)}$, where $\sigma_1$ is defined similarly. In general, $S(\gamma_1 \rVert \gamma_0)\neq S(\gamma_0 \rVert \gamma_1)$ and $\sigma_0\neq\sigma_1$, hence these two inequalities are not the same. Nonetheless, (\ref{eq:errsum}) is perhaps more useful practically since it is usually desirable to control $\alpha$ at a low level, resulting in a low $\sigma_0$. 

Fifth, from (\ref{eq:subGQFR}), there is also an upper bound $\alpha + \beta \leq  1 + \sigma_{0}\sqrt{2nS(\gamma_1 \rVert \gamma_0)}$. However, this is not quite useful since a trivial test by always accepting $H_0$ or $H_1$ leads to $\alpha+\beta=1$. However, it might be more informative in the Bayesian setting. We discuss this in Appendix \ref{Appendix:Derivation_QFRI}.

\subsection{Calculating and upper-bounding $\sigma_0$}
We now discuss the definition of the sub-Gaussian norm $\sigma_0$, and the way to calculate or bound it. We first expand $\M_1$ as $\M_1=\sum_m \mu_m | m \rangle\langle m |$, where $\{|m\rangle\}$ is an orthonormal basis of the Hilbert space. Since $0\leq \M_1 \leq I$, we have $\mu_m\in[0,1]$ for all $m$. Note that $\alpha$ can be expressed as
\[
\alpha = \Tr(\gamma_0\M_1)=\sum_m \langle m| \gamma_0 |m\rangle \mu_m \equiv \sum_m p_m \mu_m.
\]
Similarly, we have
\begin{align*}
\Tr\left(\gamma_0e^{\xi s \M_{1c}}\right) = \sum_{m:\mu_m=0} p_m e^{-\xi s \alpha} + \sum_{m:\mu_m>0} p_m e^{\xi s (\mu_m-\alpha)},    
\end{align*}
where $\M_{1c}=\M_1-\Tr(\M_1\gamma_0)I$ is the centered version of $\M_1$ with respect to $\gamma_0$. Denote
\[
q=1-\sum_{m:\mu_m=0}p_m.
\]
We can define a random variable $M$ with distribution $P_M$ as follows: $M=0$ with probability $1-q$ and $M=\mu_m$ with probability $p_m$ for all $m$ with $\mu_m>0$. Clearly, $M$ is bounded in $[0,1]$, hence it is straightforward to know that $M$ is sub-Gaussian with norm $\sigma_{MP_M}$. We define
\[
\sigma_0 \equiv \sigma_{MP_M}.
\]
Since $M\in[0,1]$, we immediately know that $\sigma_0\leq1/2$. However, this constant upper bound can be further improved. We can either numerically find $\sigma_0$ or establish a more informative upper bound. 

First, in order to calculate $\sigma_0$, note that  $\E_{M\sim P_M}M=\alpha$, and we may define $K(t)\equiv \ln\E_{M\sim P_M}e^{t (M-\alpha)}$ for $t\in\mathbb{R}$. Then, by the definition of a sub-Gaussian random variable (\ref{eq:subG}), one can calculate $\sigma_0$ by solving  a set of equations \cite{ESAIM}
\[
\begin{cases}
K(t) = \frac{1}{2}\sigma^2t^2, \\
\frac{dK(t)}{dt} = \sigma^2 t.
\end{cases}
\]
The minimal solution of $\sigma$ (if there exist more than one solution) gives $\sigma_0$. The underlying idea is that by decreasing $\sigma$ to the critical $\sigma_0$, $\sigma^2t^2/2$ must intersect $K(t)$ at some $t^\ast$ (which gives rise to the first equation), and this is achieved when they are tangent to each other at $t^\ast$ (which gives rise to the second equation). (Note that the trivial case $t=0$ should be excluded as $K(0)=\dot K(0)=0$ irrespective of $\sigma$.) 

Next, perhaps more interestingly, we show that there is a nontrivial upper bound for $\sigma_0$ that is more informative than $1/2$. To this end, we first assume $\M_1$ is a projector, where we can find the corresponding $\sigma_0$ explicitly, in terms of $\alpha$. Then we prove that it serves as an upper bound for $\sigma_0$ when $\mathbb{M}$ is a general POVM. 

Suppose $\M_1$ is a projector. The eigenvalues of $\M_1$ are in $\{0,1\}$. As a consequence, $M$ is a Bernoulli random variable with mean 
\[
\alpha=\sum_{m:\mu_m>0}\mu_mp_m=\sum_{m:\mu_m=1}p_m,
\]
and therefore have
\[
\Tr\left(\gamma_0e^{\xi s \M_{1c}}\right) =(1-\alpha)e^{-\xi s \alpha} + \alpha e^{\xi s (1-\alpha)}\leq e^{\sigma_0^2s^2/2}.
\]
For $\alpha\in(0,1)$ and $\alpha\neq0.5$, it is known that $\sigma_0=\sqrt{\frac{\alpha-0.5}{\ln(\alpha/(1-\alpha))}}$ \cite{subGnorm}, which is strictly less than 0.5. Only when $\alpha=0.5$, we have $\sigma_0=0.5$, and this is the only situation that our bound (\ref{eq:errsum}) reduces to the Pinsker bound (\ref{eq:QPinsker}). It is worth noting that when $\M_1$ is a projector, $\sigma_0$ surprisingly only depends on the false positive rate $\alpha$ without resorting to the details of the measurement plan. This is because the information of the eigenvalues of $\M_1$ has been implicitly encoded in $\alpha$. If the measurement plan is designed to control the false positive rate at a low level $\alpha$, then we immediately know the resulting $\sigma_0$ and the corresponding lower bound for the sum of error rates. In particular, in the extreme case that $\alpha\ll0.5$, we approximately have $\sigma_0 \approx \sqrt{-\frac{1}{2\ln \alpha}}$, which indicates in such an extreme situation that $\alpha$ is greatly suppressed, our bound is nontrivial (greater than 0) as long as $nS(\gamma_1\rVert\gamma_0)\lesssim \ln\left(\frac{1}{\alpha}\right)$. As a comparison, the bound in (\ref{eq:QPinsker}) is nontrivial when $nS(\gamma_1\rVert\gamma_0)\leq 2$. Our result works in a much wider range when $\alpha\ll 0.5$, which is desired experimentally. Moreover, one can impose an experimentally accessible dependence of $\alpha$ on $n$ if necessary, and our bound can still be informative in this case.

In general, $\M_1$ is not necessarily a projector, and the explicit form of $\sigma_0$ is usually unknown. Nonetheless, it is upper bounded by the the sub-Gaussian norm of a projector under the same $H_0$ and with the same false positive rate $\alpha$. We prove this in Appendix \ref{Appendix:upper_bound_sigma0}. Intuitively, this result can be understood from the fact that for all random variables bounded in $[0,1]$ and with the same mean, the Bernoulli random variable spreads out the most. Recall that the squared sub-Gaussian norm $\sigma_0^2$ is also called the variance proxy. Hence the result (\ref{eq:sigma0}) may not be too surprising. 

\subsection{Illustrative example}

\begin{figure}[htbp]  
	\centering
	\includegraphics[width=0.5\textwidth]{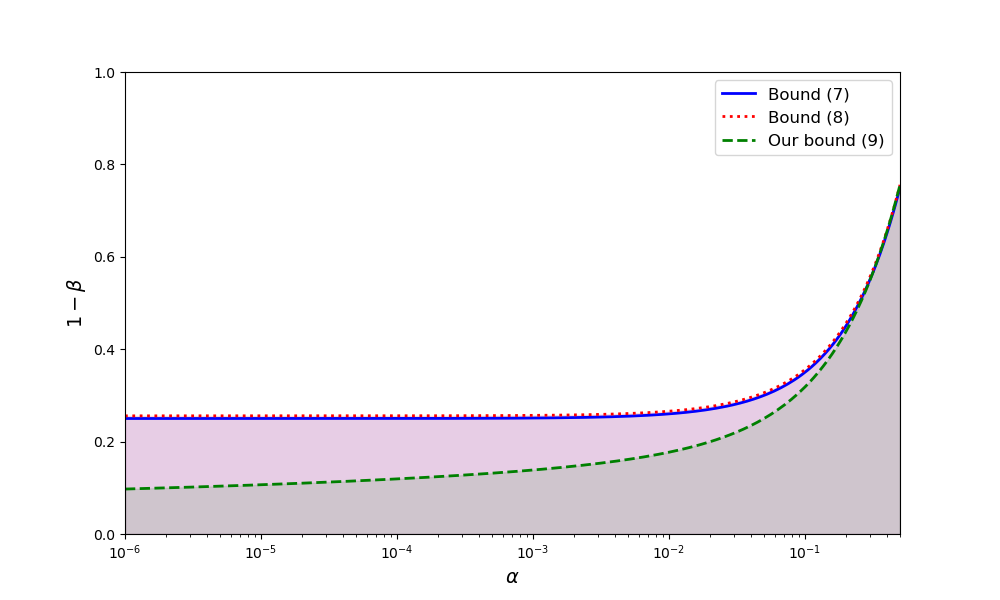}
	\caption{
			Comparison of three bounds (\ref{eq:err_bound_opt}), (\ref{eq:QPinsker}), and ours (\ref{eq:errsum}) for $m=2$ and $n=1$ in the illustrative example. These lower bounds for $\alpha+\beta$ are equivalent to upper bounds for $1-\beta$ at given $\alpha$. Recall that a smaller upper bound is more informative. The Pinsker bound (\ref{eq:QPinsker}) is always weaker than (\ref{eq:err_bound_opt}), but in this specific situation their difference is negligible. Our bound is evidently the most informative when $\alpha$ is small. In fact, since $\sigma_0\to 0$ as $\alpha\to 0$, our bound implies that $1-\beta\to 0$ as $\alpha\to 0$. Other two bounds converge to constant positive values as $\alpha\to 0$, and thus are less informative than ours.
	}
	\label{fig:three_bounds}
\end{figure}

Let us consider an illustrative example, where $\gamma_0$ and $\gamma_1$ commute. In this case, the quantum hypothesis testing problem can be reduced to a classical one, making it easier to compare different error bounds. Suppose the system we are interested in is in a state that can be described by either $\gamma_0= \frac{1}{2}|0^{\otimes m}\rangle\langle 0^{\otimes m}| + \frac{1}{2}|1^{\otimes m}\rangle\langle 1^{\otimes m}|$ or $\gamma_1= \frac{m-1}{2m}|0^{\otimes m}\rangle\langle 0^{\otimes m}| + \frac{m+1}{2m}|1^{\otimes m}\rangle\langle 1^{\otimes m}|$, where $|i^{\otimes m}\rangle \equiv (|i\rangle)^{\otimes m}$ for $i=0,1$ and $m>1$.
Then we have for all $m$ that
\begin{align*}
 S(\gamma_1\rVert\gamma_0) ={}& \frac{m-1}{2m}\ln\left(\frac{m-1}{2m}\right) + \frac{m+1}{2m}\ln\left(\frac{m+1}{2m}\right)\\ 
 & - \frac{m-1}{2m}\ln\left(\frac{1}{2}\right) -\frac{m+1}{2m}\ln\left(\frac{1}{2}\right)\\
\leq{}& \frac{1}{m^2}.
\end{align*}
When $m\gg 1$, the difference between $\gamma_0$ and $\gamma_1$ is small and $S(\gamma_1\rVert\gamma_0)\approx m^{-2}/2$. If a projective measurement is performed to $n$ independent copies of the system, then our result (\ref{eq:errsum}) states that 
\[
\alpha+\beta \geq 1-\sqrt{\frac{\alpha-0.5}{\ln(\alpha/(1-\alpha))}\frac{2n}{m^2}}.
\]
The upper bound in (\ref{eq:err_bound_opt}) can be numerically evaluated but a clean expression is not easy to obtain for $n>1$. As discussed above, our bound is always stronger than (\ref{eq:QPinsker}), which leads to $1-\sqrt{\frac{n}{m^2}}$. If we further control the type I error rate $\Tr(\gamma_0^{\otimes n}\M_1)=\alpha\ll 0.5$ by some $\M_1$, then, to leading term, our result indicates that 
\(
\beta \gtrsim 1-\sqrt{-\frac{n}{(\ln\alpha)m^2}},
\)
which gives insight into how the type II error rate $\beta$ depends on the type I error rate $\alpha$, the number of copies of the system $n$, and the dimension of the system $m$. If $\alpha$ is allowed to depend on $n$, for example, $\alpha\sim e^{-\lambda n/m^2}$ for some constant $\lambda>0$, then $\beta\gtrsim 1-1/\sqrt{\lambda}$.

In Fig.~\ref{fig:three_bounds}, we plot different error bounds for $m=2$ and $n=1$. In such a simple setting, each bound can be precisely evaluated for a straightforward comparison. For any given $\alpha$, these lower bounds for the sum of error rates can be rewritten as upper bounds for $1-\beta$, respectively. Our bound (\ref{eq:errsum}) is evidently more informative than the other two bounds when $\alpha$ is controlled at a small level.

\section{More applications}\label{sec:more_appl}
Due to their generality, our results provide a natural starting point for a variety of problems, including but not limited to quantum hypothesis testing. We outline several potential applications in physics and machine learning below.

\subsection{Thermodynamic hypothesis testing}
It turns out that a classical counterpart of (\ref{eq:errsum}) can provide insights to thermodynamic inference \cite{ARCMP:Seifert} via hypothesis testing. The classical error bound was established in our previous work \cite{Me:ISIT} and can also be easily obtained by assuming all involved operators are commutative in Appendix \ref{Appendix:Derivation_QFRI}. Thus taking the trace is replaced with taking the expectation, and the quantum relative entropy is replaced with the Kullback-Leibler divergence. The inequality now reads 
\begin{align}\label{eq:classical_bound}
\alpha+\beta\geq 1-\sigma_0\sqrt{2n\KL(P_1\rVert P_0)}. 
\end{align}

In the classical case, the probabilities can be associated with a forward/backward process in the context of stochastic thermodynamics \cite{RPP:Seifert}. Concretely, we consider a nonequilibrium steady state. Let $\omega=(X_0,\ldots,X_T)$ be a stochastic trajectory of a Brownian particle in a time interval $[0,T]$. Let $\omega^\dagger=(X_T,\ldots,X_0)$ be the time-reversed trajectory. One can establish a one-to-one map $\mathcal{I}$ between $\omega$ and $\omega^\dagger$, say, $\omega^\dagger = \mathcal{I}(\omega)$. Suppose $P(\omega)$ is the probability measure for $\omega$. Then we have for Markovian dynamics that 
\(
	P(\omega) = \prod_{t=1}^{T} p(X_{t}|X_{t-1}) \times p_0(X_0),
\)
where $p_0$ denotes the marginal distribution of $X_0$ and $p(X_{t}|X_{t-1})$ denotes the conditional distribution of $X_t$ given $X_{t-1}$. The map $\mathcal{I}$ induces the pushforward measure $\mathcal{I}_{\#}P$ such that 
\(
\mathcal{I}_{\#}P(\omega) \equiv P^\dagger(\omega^\dagger),
\)
where $P^\dagger(\omega^\dagger)$ is given by
\(
	P^\dagger(\omega^\dagger) = \prod_{t=T}^{1} p(X_{t-1}|X_{t}) \times p_T(X_T). 
\)
The KL divergence between $P(\omega)$ and $P^\dagger(\omega^\dagger)$ is thus
\begin{align*}
	&\KL(P(\omega) \rVert P^\dagger(\omega^\dagger))\\ ={}&\sum_{X_0,\ldots,X_T} P(X_0,\ldots,X_T) \times \sum_{t=1}^{T-1}\ln\left(\frac{p(X_t|X_{t-1})}{p(X_{t-1}|X_t)}\right)\\
	&{} \qquad \qquad + \sum_{X_0,\ldots,X_T} P(X_0,\ldots,X_T)\times\ln\left(\frac{p_0(X_0)}{p_T(X_T)}\right).
\end{align*}
In a nonequilibrium steady state, the marginals $p_0$ and $p_T$ are the same distribution $\pi$, and
\begin{align*}
&\sum_{X_0,\ldots,X_T} P(X_0,\ldots,X_T)\ln(p_0(X_0)/p_T(X_T))\\
={}& \sum_{X_0} \pi(X_0) \ln(\pi(X_0)) - \sum_{X_T} \pi(X_T) \ln(\pi(X_T)) = 0.
\end{align*}
Therefore, $\KL(P(\omega) \rVert P^\dagger(\omega^\dagger))$ is only determined by the average asymmetry in the transition dynamics in forward and backward processes, which is known to be the entropy production $\Delta S$ on the trajectory level. We thus have
\[
\Delta S = \KL(P(\omega)\rVert P^\dagger(\omega^\dagger)) = \KL(P(\omega)\rVert \mathcal{I}_{\#}P(\omega)).
\]
Given $n$ observations of trajectory data $\{\omega_i\}_{i=1}^n$, one may want to determine the arrow of time by testing $H_0: \omega_i\overset{iid}{\sim} I_\# P$ vs. $H_1:\omega_i\overset{iid}{\sim} P$ based on the empirical distribution of $\omega_i$ (or equivalently, of some summary statistic of $\omega_i$). For any test function $\Phi$, taking values in $[0,1]$ and with error rates $\alpha$ and $\beta$, the bound (\ref{eq:classical_bound}) becomes 
\begin{align}\label{eq:thermodynamic_bound}
\alpha+\beta\geq 1-\sigma_0\sqrt{2n\Delta S},
\end{align}
where $\sigma_{0}$ is the sub-Gaussian norm of $\Phi$ under $H_0$. Since $\Phi$ is in $[0,1]$, and thus sub-Gaussian, $\sigma_0$ is always no greater than 0.5. So we can conclude that $\alpha+\beta\geq 1-\sqrt{n\Delta S/2}$ even without knowing the details of $\Phi$. This result reflects the intrinsic difficulty in identifying time's arrow near equilibrium where $\Delta S\approx 0$ and $\alpha+\beta \gtrsim 1$. Any test function used in the hypothesis testing is no better than random guess (i.e., always accepting $H_0$ with probability 0.5), where $\alpha+\beta =1$. So, it is practically impossible to tell the arrow of time unless the size of the trajectory dataset $n$ is at least on the order of $(\Delta S)^{-1}$. In contrast to existing work on estimating the entropy product quantitatively \cite{PRE:Dechant,PRE:Hasegawa}, our result addresses the problem of thermodynamic inference in a more qualitative way.

\subsection{Speed limit}
Let $\OP$ be an arbitrary observable of interest, $\Ham$ be the Hamiltonian, $\gamma_0 = \gamma_t$ be the density operator of a system at time $t$, and $\gamma_1=\gamma_{t+dt}$. Our result (\ref{eq:subGQFR}) implies that
\[
|\Tr(\gamma_{t+dt}\OP) -\Tr(\gamma_t\OP)|\leq \sigma_{Ot}\sqrt{2S(\gamma_{t+dt}\rVert \gamma_t)},
\]
where $\sigma_{Ot}$ is the sub-Gaussian norm of $\OP$ under $\gamma_t$. Note that this result does \emph{not} assume that the dynamics of $\gamma_t$ is differentiable. But if we assume that the system undergoes unitary dynamics, then, to second order,
\begin{align}\label{eq:S_in_speedlimit}
S(\gamma_{t+dt}\Vert\gamma_{t})\approx \frac{dt^2}{2}\Tr([\Ham,[\Ham,\gamma_t]]\ln\gamma_t),
\end{align}
where we take Planck's constant $\hbar$ to be $1$. Details to derive \eqref{eq:S_in_speedlimit} are given in Appendix~\ref{Appendix:Speed limit}. Denote $\langle \OP \rangle_{t+dt}=\Tr(\gamma_{t+dt}\OP)$ and $\langle \OP \rangle_{t}=\Tr(\gamma_{t}\OP)$. Then the speed of $\Tr(\gamma_t\OP)$ is
\[
\langle \dot \OP \rangle_t \equiv \frac{d\langle \OP \rangle_{t}}{dt} \approx \frac{\langle \OP \rangle_{t+dt} - \langle \OP \rangle_{t}}{dt}.
\]
Combining (\ref{eq:subGQFR}) and \eqref{eq:S_in_speedlimit} yields that
\begin{align*}
|\langle \dot \OP \rangle_t| & \lesssim \sigma_{Ot}\sqrt{2\frac{d^2 S(\gamma_{t+dt}\rVert\gamma_t)}{dt^2}}\\
&\lesssim \sigma_{Ot}\sqrt{\Tr([\Ham,[\Ham,\gamma_t]]\ln\gamma_t)}.
\end{align*}
This result is different from existing speed limits such as the Mandelstam-Tamm bound \cite{MT_bound}, which states that 
\[
|\langle \dot \OP \rangle_t|\leq 2(\Delta_t\OP)(\Delta_t\Ham),
\] 
where $\Delta_t\OP=\sqrt{\Tr(\OP^2\gamma_t)-\Tr(\OP\gamma_t)^2}$ and $\Delta_t\Ham=\sqrt{\Tr(\Ham^2\gamma_t)-\Tr(\Ham\gamma_t)^2}$ are the standard deviations of $\OP$ and $\Ham$ with respect to $\gamma_t$, respectively.

As an illustrative example to show the difference between our bound and the Mandelstam-Tamm bound, let us consider a two-level system with $\Ham=n_1|n_1\rangle\langle n_1| + n_2|n_2\rangle\langle n_2|$ and $\gamma_t=\frac{1}{2}|n_1\rangle\langle n_1|+\frac{1}{2}|n_2\rangle\langle n_2|$. It is straightforward to check that $[\gamma_t,\Ham]=0$, and our bound suggests $\langle\dot\OP\rangle_t=0$. On the other hand, $\Tr (\Ham^2\gamma_t)=\frac{1}{2}(n_1^2+n_2^2)$ and $\Tr(\Ham\gamma_t)=\frac{1}{2}(n_1+n_2)$, and thus $\Delta_t\Ham=\frac{1}{2}|n_1-n_2|$, which is greater than $0$ as long as $n_1\neq n_2$. Since $\Delta_t\OP$ is also greater than $0$ in general, the Mandelstam-Tamm bound yields a nonzero upper bound and is therefore less informative than our result. There are other speed limit bounds (under different settings) \cite{PRX:SL}, and it will be interesting to systematically compare our result with them in various physical systems in the future. 

\subsection{Potential application to generalization analysis}
As discussed in Sec.~\ref{sec:Intro}, it is of great interest to understand the difference between population and empirical risks for a learning algorithm. In particular, a core quantity is $\sup_{h}|\E_{P_Z}L(h,Z)-\E_{P_n}L(h,Z)|$, where the supremum is over a given function class that $h$ lies in. This function class is determined by the learning algorithm in question. For example, linear regression corresponds to a class of linear functions. To bound this difference, a standard approach is to use the Rademacher complexity \cite{AoS:Bartlett}. If $L(h,Z)$ is sub-Gaussian (which is always satisfied by a bounded loss function), then our work provides an alternative upper bound:
\begin{align*}
&\E\left[\sup_{h}|\E_{P_Z}L(h,Z)-\E_{P_n}L(h,Z)|\right]\\ 
\leq{}& \E\left[\sup_g \sigma_{g}\sqrt{2\KL(g_{\#}P_n\rVert g_{\#}P_Z)}\right]\\
\leq{}& \left(\sup_g \sigma_g\right) \times \E\left[\sup_g \sqrt{2\KL(g_{\#}P_n\rVert g_{\#}P_Z)}\right],
\end{align*}
where $\E[\cdot]$ denotes the expectation with respect to the training data. We define $g(Z)$ as $g(Z)\equiv L(h,Z)$, and $g_\#P_n$ or $g_\#P_Z$ is the corresponding pushforward probability measure induced by $g$. The sub-Gaussian norm of $g(Z)$ under $g_\#P_Z$ is denoted as $\sigma_g$. For this bound to be nontrivial, we need $\KL(g_{\#}P_n\rVert g_{\#}P_Z)<\infty$ for all $g$. This condition is generally satisfied in classification tasks with a finite number of labels. For regression problems, however, the learning algorithm typically needs to incorporate (kernel) smoothing of the observed data to ensure finiteness of the relative entropy. While a comprehensive analysis of this issue is left for future work, we briefly illustrate the applicability of the bound through a simple yet concrete example.

Consider a two-class classification problem with the 0-1 loss. We have $\sigma_g \leq \sqrt{\frac{p_g-0.5}{\ln(p_g/(1-p_g))}}$ by noting the similarity between this problem and the hypothesis testing problem studied in this work, where $p_g = \E_{g_\#P_Z}[g(Z)]\in[0,1]$. Suppose there are $K$ models that correspond to $g_1,\ldots,g_K$. Then this bound reveals that a model with the minimal empirical risk will also have a small population risk if each model is better than random guess (i.e., $\sigma_{g_k}<0.5$ for all $g_k$), and $\KL(g_{k\#}P_n\rVert g_{k\#}P_Z)$ is small for all $g_k$ (i.e., the empirical distribution of $g_k(Z_i)$ for training data $Z_i$ is close to its population distribution). This characterization of generalization in terms of the similarity between the empirical and population distributions is qualitatively different from classical generalization bounds based on the complexity of the function class \cite{AoS:Bartlett}. We leave further analysis inspired by this result for future study.

\section{Conclusion and Outlook}\label{sec:conclusion}
In this work, we have established the quantum fluctuation-response inequality (\ref{eq:QFR}) by nontrivially generalizing the classical result to the quantum setting, which provides a generic upper bound on the difference between expectation values of a physical observable in two states, in terms of the quantum relative entropy between the two states. We further introduce its sub-Gaussian version (\ref{eq:subGQFR}), which is particularly relevant for bounded operators in real physical systems. Based on this result, we analyze quantum hypothesis testing and derive a non-asymptotic error bound (\ref{eq:errsum}) that complements, and in some cases is more informative than, existing bounds. Our bound separates the intrinsic distinguishability of the two states, measured by the quantum relative entropy, from the sensitivity of the particular
measurement used to distinguish them, and is thus measurement dependent. We also explore the broader implications of our findings in both quantum and classical settings, including applications to thermodynamic hypothesis testing in stochastic thermodynamics, quantum speed limits in statistical physics, and the generalization bound for empirical risk minimizers in machine learning.

Several promising directions for future research remain to be explored. First, the central message of this work is that the response of an observable is governed jointly by the distance between the reference and perturbed states and by how strongly the chosen observable fluctuates in the reference state. Quantum relative entropy and sub-Gaussian concentration provide a concise way to quantify these two effects.  We expect that this viewpoint will serve as a useful basis for further developments in quantum nonlinear response theory.

Second, the quantum hypothesis testing problem considered in this work is static (or offline) in the sense that the number of copies of the system $n$ is fixed in advance. Developing an analogous bound for the dynamic (or online) setting, in which copies of the system are obtained sequentially, is another promising direction for future investigation.

Third, for the speed-limit problem under unitary dynamics, a natural extension is to derive a finite-time bound on the change in the expectation value,
\begin{align*}
	|\langle \OP \rangle_{t+T}-\langle\OP\rangle_t|
	&\le \int_0^T \left|\langle \dot{\OP}\rangle_{t+\tau}\right|d\tau\\
	\lesssim \int_0^T & \sigma_{O_{t+\tau}}
	\sqrt{\Tr([\Ham,[\Ham,\gamma_{t+\tau}]]\ln\gamma_{t+\tau})}d\tau,
\end{align*}
which is closely connected to dynamical response theory. It would be interesting to investigate what new physical insights can be gained from this perspective within our framework. In particular, such an extension may help elucidate the physical meaning of the sub-Gaussian norm $\sigma_{\OP_t}$.
  
Fourth, as discussed above, our work may also offer a new perspective on the generalization bounds of empirical risk minimizers in machine learning. An interesting direction for future research is to apply our bound to concrete learning problems, thereby providing a clearer comparison with existing bounds.

Finally, we hope that these directions will stimulate further research at the intersection of statistical physics, quantum information, and machine learning.

\section*{Acknowledgments}
Part of the work was completed while Y.W. was a graduate student in the Department of Statistics at Iowa State University. Y.W. gratefully thank Prof. Dan Nettleton and Prof. Huaiqing Wu for helpful discussions, and Prof. Thomas Iadecola for reviewing an earlier version of this work. Y.W. was supported by the US National Science Foundation under grant HDR: TRIPODS 19-34884.


\appendix

\section{Derivation of the quantum fluctuation-response inequality}\label{Appendix:Derivation_QFRI}
Rather than directly prove the QFRI and its sub-Gaussian version, we start from a more general setting by taking the Bayesian case into account and get back to the main results as special cases. The following two inequalities play an important role in our proof. 
\begin{lem}[Klein's inequality]
Let $A$ and $B$ be Hermitian matrices of the same size, and $f$ be a differentiable convex function defined on the union of the supports of $A$ and $B$. Then we have
\begin{align}\label{eq:Klein}
\Tr[f(B)] \geq \Tr[f(A)] + \Tr[f'(A)(B-A)].
\end{align}
\end{lem}

\begin{lem}[The Golden-Thompson inequality] 
Let $A$ and $B$ be Hermitian matrices, then
\begin{align}\label{eq:GT}
\Tr (e^{A+B}) \leq \Tr(e^A e^B).
\end{align}
\end{lem}
In the following, we let $\pi_0$ and $\pi_1$ be two positive real numbers. For a generic Hermitian operator $\OP$, two density operators $\gamma_0$, $\gamma_1$, and $s>0$, we define 
\begin{align}
B &= \xi s \left[\pi_1\OP - \pi_0\Tr(\gamma_0\OP)I\right] + \ln\gamma_0,\label{eq:newB}\\
A &= \ln\gamma_1 + \Tr(\gamma_1 B-\gamma_1\ln\gamma_1)I,\label{eq:newA}
\end{align}
where \[
\xi = \text{sgn}(\pi_1\Tr(\gamma_1\OP) - \pi_0\Tr(\gamma_0\OP)).
\]
Note that $\Tr(\gamma_1I)=\Tr(\gamma_1)=1$. It is straightforward to verify that
\begin{align}\label{eq:BminusA}
&{}\Tr(\gamma_1(B-A)) \notag\\
={}& \Tr(\gamma_1B) - \Tr(\gamma_1\ln\gamma_1)- \Tr(\gamma_1B-\gamma_1\ln\gamma_1)\Tr(\gamma_1I)\notag\\
= {}& 0.
\end{align}
We further denote $\lambda = \Tr(\gamma_1 B-\gamma_1\ln\gamma_1)$. So 
\[
A = \ln\gamma_1 + \lambda I.
\]
Now, we let $f(x)=e^x$ in Klein's inequality. Then $f'(x)=e^x$, and we have
\begin{align}
& \Tr(e^B)\notag\\
\geq{}& \Tr(e^A) + \Tr(e^A(B-A))\notag\\
={}& \Tr\left[\exp( \ln\gamma_1 + \lambda I )\right] + \Tr\left[\exp(\ln\gamma_1+\lambda I)(B-A)\right]\notag\\ 
&{} \qquad \text{[inserting Eq. (\ref{eq:newA})]}\notag\\
={}& \Tr(\gamma_1e^{\lambda I}) + \Tr(\gamma_1 e^{\lambda I}(B-A)) \quad \text{(since $[\ln\gamma_1,I]=0$)}\notag\\
={}&  \Tr(\gamma_1e^\lambda I) + \Tr(\gamma_1e^\lambda I(B-A))  \quad \text{(since $e^{\lambda I}=e^\lambda I$)}\notag\\
={}& e^\lambda \Tr(\gamma_1) + e^\lambda\Tr\left(\gamma_1(B-A)\right)\notag\\
={}& e^\lambda, \qquad \text{[by Eq. (\ref{eq:BminusA}) and $\Tr(\gamma_1)=1$]}\notag
\end{align}
by which we immediately obtain
\begin{align}\label{eq:RHS}
{}&\ln\Tr(e^B) 
\geq \lambda = \Tr(\gamma_1 B-\gamma_1\ln\gamma_1)\notag\\
={}& \Tr\left(\gamma_1\xi s \left[\pi_1\OP - \pi_0\Tr(\gamma_0\OP)I\right] + \ln\gamma_0 \right)\notag\\
&{} -\Tr(\gamma_1\ln\gamma_1) \qquad\text{[inserting Eq. (\ref{eq:newB})]}\notag\\
={}& \xi s [\pi_1\Tr(\gamma_1\OP) -\pi_0\Tr(\gamma_0\OP)\Tr(\gamma_1I)] \notag\\
&{} + \Tr(\gamma_1\ln\gamma_0) -\Tr(\gamma_1\ln\gamma_1) \notag\\
={}& \xi s  [\pi_1\Tr(\gamma_1\OP) -\pi_0\Tr(\gamma_0\OP)] - \QKL\notag\\
{}& \qquad\text{[by the definition of $\QKL$ and $\Tr(\gamma_1)=1$]}\notag\\
={}& s  |\pi_1\Tr(\gamma_1\OP) -\pi_0\Tr(\gamma_0\OP)| - \QKL.\notag\\
{}&\qquad \text{[by the definition of $\xi$]}
\end{align}
On the other hand, by (\ref{eq:newB}), we note that
\begin{align}\label{eq:LHS}
&{}\ln\Tr(e^B)  
= \ln\Tr\left(\exp( \xi s \left[\pi_1\OP - \pi_0\Tr(\gamma_0\OP)I\right] + \ln\gamma_0)\right)\notag\\
\leq{}& \ln\Tr\left(\gamma_0 \exp( \xi s \left[\pi_1\OP - \pi_0\Tr(\gamma_0\OP)I\right])\right) \notag\\
{}& \qquad\text{(by Golden-Thompson)}\notag\\
={}& \ln\Tr\left(\gamma_0 \exp(\xi s[\pi_1\OP - \pi_0\Tr(\gamma_0\OP)I \right. \notag\\
{}&\qquad  \qquad \qquad \qquad \left.
- \pi_1\Tr(\gamma_0\OP)I + \pi_1\Tr(\gamma_0\OP)I])\right)\notag\\
={}& \ln\Tr\left(\gamma_0 e^{\xi s[\pi_1\OP_c + (\pi_1-\pi_0)\Tr(\gamma_0\OP)I]}\right)\notag\\
{}& \qquad\text{[Note that $\OP_c=\OP - \Tr(\gamma_0\OP)I]$}\notag\\
={}& \ln\Tr\left(\gamma_0 e^{\xi s \pi_1 \OP_c}e^{\xi s (\pi_1-\pi_0)\Tr(\gamma_0\OP)I}\right) \quad\text{(since $[\OP_c,I]=0$)}\notag\\
={}& \ln\Tr\left(\gamma_0 e^{\xi s \pi_1 \OP_c}e^{\xi s (\pi_1-\pi_0)\Tr(\gamma_0\OP)}I\right) \quad\text{(since $e^{cI}=e^c I$)}\notag\\
={}& \ln\left\{ e^{\xi s (\pi_1-\pi_0)\Tr(\gamma_0\OP)} \Tr\left(\gamma_0e^{\xi s \pi_1 \OP_c}\right)\right\}\notag\\
={}&\ln\Tr\left(\gamma_0e^{\xi s \pi_1 \OP_c}\right) + \xi s (\pi_1-\pi_0)\Tr(\gamma_0\OP).
\end{align}
Combining (\ref{eq:RHS}) and (\ref{eq:LHS}), and noting $s>0$, we have
\begin{align}\label{eq:sineq}
{}&|\pi_1\Tr(\gamma_1\OP) -\pi_0\Tr(\gamma_0\OP)|\notag\\ 
\leq{}& \frac{1}{s}\ln\Tr\left(\gamma_0e^{\xi s \pi_1 \OP_c}\right) + \xi (\pi_1-\pi_0)\Tr(\gamma_0\OP) + \frac{1}{s}\QKL.
\end{align}
For typical operators of interest, (\ref{eq:sineq}) holds for all $s>0$ where each term is well defined. Hence we have that
\begin{align}\label{eq:GQFR}
&{}|\pi_1\Tr(\gamma_1\OP) -\pi_0\Tr(\gamma_0\OP)|\notag\\
\leq{}& \inf_{s>0} \left[ \frac{1}{s}\ln\Tr\left(\gamma_0e^{\xi s \pi_1 \OP_c}\right) \right.\notag\\
&{} \qquad \left. + \xi (\pi_1-\pi_0)\Tr(\gamma_0\OP) + \frac{1}{s}\QKL\right].
\end{align}
Using an orthonormal basis $\{|i\rangle\}$ that diagonalizes $\OP_c$, we further have  
\begin{align}
\Tr\left(\gamma_0e^{\xi s \pi_1 \OP_c}\right) &{} = \Tr\left(\gamma_0\sum_{i}|i\rangle\langle i|e^{\xi s \pi_1 \OP_c} \right)\notag\\
&{}=\sum_i\langle i|\gamma_0|i\rangle e^{\xi s\pi_1o_i}\equiv\sum_{i}p_ie^{\xi s\pi_1o_i},\notag
\end{align}
where $\{o_i\}$ are eigenvalues of $\OP_c$. This defines a random variable $O_c$ that equals $o_i$ with probability $p_i$. Also, $O_c$ has mean zero under $P_0\equiv\{p_i\}$ since 
\[
\E_{O\sim P_0}O=\Tr(\gamma_0\OP_c)=\Tr(\gamma_0\OP)-\Tr(\gamma_0\OP)\Tr(\gamma_0I)=0.
\]
If $O_c$ is sub-Gaussian under $P_0$, then $\OP_c$ may also be said to be sub-Gaussian under $\gamma_0$ in the following sense: by (\ref{eq:subG}), we have
\begin{align}
\Tr\left(\gamma_0e^{\xi s \pi_1 \OP_c}\right) &{}= \E_{O_c\sim P_0}e^{\xi s\pi_1O_c}\notag\\
&{} \leq e^{\sigma_{OP_0}^2(\xi s\pi_1)^2/2} = e^{(\pi_1\sigma_{OP_0})^2s^2/2},\label{eq:GsubG}
\end{align} 
where $\sigma_{OP_0}$ is the sub-Gaussian norm of $O_c$ under $P_0$. Hence (\ref{eq:GQFR}) can be further written as
\begin{align}\label{eq:GsubGQFR}
&{}|\pi_1\Tr(\gamma_1\OP) -\pi_0\Tr(\gamma_0\OP)| \notag\\
\leq{}&\inf_{s>0} \left[ \frac{1}{s}\ln\Tr\left(\gamma_0e^{\xi s \pi_1 \OP_c}\right) + \xi (\pi_1-\pi_0)\Tr(\gamma_0\OP)\right.\notag\\
&\qquad \qquad + \left. \frac{1}{s}\QKL\right]\notag\\
\leq{}& \inf_{s>0} \left[ \frac{1}{2s}(\pi_1\sigma_{OP_0})^2s^2 + \xi (\pi_1-\pi_0)\Tr(\gamma_0\OP)\right.\notag\\
&\qquad \qquad + \left. \frac{1}{s}\QKL\right]\notag\\
={}& \inf_{s>0} \left[ \frac{1}{2}(\pi_1\sigma_{OP_0})^2s + \xi (\pi_1-\pi_0)\Tr(\gamma_0\OP) + \frac{1}{s}\QKL\right]\notag\\
={}&\pi_1\sigma_{OP_0}\sqrt{2\QKL} + \xi (\pi_1-\pi_0)\Tr(\gamma_0\OP),
\end{align} 
where the infimum is achieved at $s=\sqrt{2\QKL}/\pi_1\sigma_{OP_0}$. 

Now we consider two special cases of (\ref{eq:GsubGQFR}). First, $\pi_0=\pi_1=1$. In this case, the sub-Gaussian quantum fluctuation-response inequality (\ref{eq:subGQFR}) is recovered. Second, $\pi_0,\pi_1\in(0,1)$ and $\pi_0+\pi_1=1$. Let
$\OP=\M_1$ as an element of a POVM. Note that the type I error rate $\alpha=\Tr(\gamma_0\M_1)$ and the type II error rate $\beta = \Tr(\gamma_1\M_0) = 1-\Tr(\gamma_1\M_1)$. We then have
\begin{align*}
|\pi_1(1-\beta)-\pi_0\alpha| \leq \pi_1\sigma_{OP_0}\sqrt{2\QKL} + \xi (\pi_1-\pi_0)\alpha,
\end{align*}
which leads to
\begin{align}
 &\pi_1\left[1-\sigma_{OP_0}\sqrt{2\QKL}\right] - \xi(\pi_1-\pi_0)\alpha \notag\\
 \leq{}& \pi_0\alpha + \pi_1\beta \notag\\
 \leq{}&\pi_1\left[1+\sigma_{OP_0}\sqrt{2\QKL}\right] + \xi(\pi_1-\pi_0)\alpha.
\end{align}
When $\pi_0=\pi_1=1/2$, let $\sigma_0\equiv\sigma_{OP_0}$, then our error bound (\ref{eq:errsum}) is recovered. In general, however, this result depends on $\xi$. In the case that $\xi=1$, we have $$\pi_1\Tr(\gamma_1\M_1) > \pi_0\Tr(\gamma_0\M_1)\Longrightarrow \pi_0\alpha + \pi_1\beta < \pi_1.$$ Under such a constraint, we obtain the following lower and upper bounds
\begin{align*}
&{}\alpha+\beta \geq 1-\sigma_{OP_0}\sqrt{2\QKL}  \text{\ and}\notag\\
&{}\left(\frac{2\pi_0}{\pi_1}-1\right)\alpha + \beta \leq 1+\sigma_{OP_0}\sqrt{2\QKL}.
\end{align*}
Similarly, for $\xi=-1$, the constraint is $\pi_0\alpha + \pi_1\beta > \pi_1$, and we have error bounds as
\begin{align*}
&{}\left(\frac{2\pi_0}{\pi_1}-1\right)\alpha + \beta  \geq 1-\sigma_{OP_0}\sqrt{2\QKL}   \text{\ and}\notag\\
&{} \alpha+\beta \leq 1+\sigma_{OP_0}\sqrt{2\QKL}.
\end{align*}
These bounds provide new insights to the sum of error rates in the Bayesian setting.

As a byproduct of our work, the Gibbs variational principle can be obtained. Consider $\pi_0=\pi_1=1$. A closer inspection of the steps to obtain (\ref{eq:RHS}) shows that if we do not introduce $\xi$ in (\ref{eq:newB}) and simply let $B=s \left[\OP - \Tr(\gamma_0\OP)I\right] + \ln\gamma_0$, then we have for all $s>0$ that 
\begin{align*}
	&\ln\Tr(e^B) \notag\\
	= {} &\ln\Tr\left(\exp( s \left[\OP - \Tr(\gamma_0\OP)I\right] + \ln\gamma_0)\right)
	\notag\\
	\ge {}& \Tr\left(\gamma_1 s \left[\OP - \Tr(\gamma_0\OP)I\right] + \ln\gamma_0 \right)-\Tr(\gamma_1\ln\gamma_1)\notag\\
	={}& s \left(\Tr(\gamma_1\OP) - \Tr(\gamma_0\OP)\right) +\Tr(\gamma_1\ln\gamma_0) - \Tr(\gamma_1\ln\gamma_1).
\end{align*}
If we let $s=1$ and
\[
-\Ham/T = \left[\OP - \Tr(\gamma_0\OP)I\right] + \ln\gamma_0,
\]
where $\Ham$ is the Hamiltonian and $T$ is the temperature, then 
\begin{align*}
	\Tr(\gamma_1\OP) -{}& \Tr(\gamma_0\OP) \notag\\
	\le {}& \ln\Tr(e^{-\Ham/T}) +\Tr(\gamma_1\ln\gamma_1) - \Tr(\gamma_1\ln\gamma_0).
\end{align*}
Since \(
\OP = -\Ham/T-\ln\gamma_0 + \Tr(\gamma_0\OP)I
\), we further obtain
\[
\Tr(\gamma_1\OP)=-\Tr(\gamma_1\Ham/T) - \Tr(\gamma_1\ln\gamma_0) + \Tr(\gamma_0\OP).
\] 
Finally, we arrive at
\[
-\ln\Tr(e^{-\Ham/T}) \le \Tr(\gamma_1\Ham/T) + \Tr(\gamma_1\ln\gamma_1),
\]	
which is the Gibbs variational principle. Note that $\gamma_1$ is arbitrary, as long as the involved traces are well defined. As can be verified directly, equality is attained at the thermal state $\gamma_1 = e^{-\Ham/T}/\Tr(e^{-\Ham/T})$. 
Also, the index $i$ in the density operator $\gamma_i$ is a dummy label and has no physical significance.

\section{Upper bounding the sub-Gaussian norm}\label{Appendix:upper_bound_sigma0}
We will show that the sub-Gaussian norm of $\M_1$ under $H_0:\gamma\sim\gamma_0$ as defined in the main text can be upper bounded by that of a $\{0,1\}$-valued test function (a Bernoulli random variable) with the same $\alpha$. We use a set of orthonormal basis vectors $\{|m\rangle\}$ to diagonalize $\M_1$ with eigenvalues $\{\mu_m\}$, and
\begin{align*}
\alpha={}&\Tr(\gamma_0\M_1)=\Tr\left(\gamma_0\sum_{m}|m\rangle\langle m|\M_1\right)\notag\\
={}&\sum_{m}p_m\mu_m=\sum_{m:\mu_m>0} p_m\mu_m,
\end{align*}
where $p_m=\langle m|\gamma_0|m\rangle$ and $\mu_m\in[0,1]$ is the eigenvalue of $\M_1$ corresponding to $|m\rangle$. This procedure induces a random variable $M$ that takes on values in $\{\mu_m\}$ with the corresponding probability $P_M=\{p_m\}$. Evidently, $\E_{M\sim P_M}M=\alpha$ by construction. Using $M$, we have
\begin{align*}
\Tr(\gamma_0 e^{\xi s \M_1}) 
&= \E_{M\sim P_M}e^{\xi s M} =\sum_{m}p_m e^{\xi s\mu_m}\notag\\
&=\left(\sum_{m:\mu_m=0} + \sum_{m:\mu_m>0}\right) p_m e^{\xi s \mu_m}\notag\\
&=\sum_{m:\mu_m=0}p_m + \sum_{m:\mu_m>0}p_m e^{\xi s \mu_m}\notag\\
&= 1-\sum_{m:\mu_m>0}p_m + \sum_{m:\mu_m>0}p_m e^{\xi s \mu_m} \notag\\
&= 1 + \sum_{m:\mu_m>0}p_m\left( e^{\xi s \mu_m}-1\right).
\end{align*}
On the other hand, let us define a Bernoulli random variable $Y\sim P_Y=\text{Bernoulli}(\alpha)$. That is, $Y=0$ with probability $1-\alpha$ and $Y=1$ with probability $\alpha$. Note $Y$ and $M$ have the same mean $\alpha$, and
\begin{align*}
\E_{Y\sim P_Y}e^{\xi sY}
&=(1-\alpha)+\alpha e^{\xi s}=1 + \alpha\left(e^{\xi s}-1\right)\notag\\
&=1 + \left (\sum_{m:\mu_m>0} p_m\mu_m \right) \left(e^{\xi s}-1\right)\notag\\
&= 1+ \sum_{m:\mu_m>0}p_m\left(\mu_m e^{\xi s}-\mu_m\right).
\end{align*}
The difference between $\E_{Y\sim P_Y}e^{\xi sY}$ and $\E_{M\sim P_M}e^{\xi s M}$ can then be calculated as
\begin{align*}
&\E_{Y\sim P_Y}e^{\xi sY} - \E_{M\sim P_M}e^{\xi s M}\notag\\
={}& \sum_{m:\mu_m>0} p_m\left(\mu_m e^{\xi s}-\mu_m-e^{\xi s\mu_m}+1\right)\notag\\
\equiv{}& \sum_{m:\mu_m>0} p_m f(\mu_m,s),
\end{align*}
where $f(x,s)\equiv xe^{\xi s}-e^{\xi s x} + (1-x)$ for $x\in[0,1]$ and $s>0$. Note that 
\[
\frac{\partial f(x,s)}{\partial s}=\xi xe^{\xi s}-\xi x e^{\xi s x}.
\] 
If $\xi =1$, then 
\[
\frac{\partial f(x,s)}{\partial s}=xe^{s}-x e^{s x}=x(e^{s}-e^{s x})\geq 0
\] 
since $x\in[0,1]$ and thus $e^{s}\geq e^{s x}$ for $s>0$. If $\xi=-1$, then 
\[
\frac{\partial f(x,s)}{\partial s}=-xe^{-s}+x e^{-s x}=x(e^{-sx}-e^{-s})\geq0
\] 
since $e^{-sx}\geq e^{-s}$ when $x\in[0,1]$ and $s>0$. Hence we conclude that 
\[
\frac{\partial f(x,s)}{\partial s}\geq 0,
\]
which indicates that $f$ is increasing in $s$. Also note $f(x,0)=0$. We thus finally arrive at $$f(x,s)\geq0,$$ which implies that
\begin{align*}
\E_{M\sim P_M}e^{\xi s (M-\alpha)} \leq \E_{Y\sim P_Y}e^{\xi s(Y-\alpha)} \leq e^{\sigma^2_Ys^2/2},
\end{align*}
where $\sigma_Y$ is the sub-Gaussian norm of the Bernoulli random variable $Y$, which is known to be $\sigma_Y = \sqrt{(0.5-\alpha)/\ln((1-\alpha)/\alpha)}$. Since by definition, the sub-Gaussian norm of $M$ is an infimum, it holds that $\sigma_M\leq \sigma_Y$. We thus have proved that the sub-Gaussian norm of $\M_1$ under $H_0$ is upper bounded by $\sqrt{(0.5-\alpha)/\ln((1-\alpha)/\alpha)}$.

\section{Speed limit results}\label{Appendix:Speed limit}
In the speed limit example, we assume $\gamma_t$ is positive definite and under unitary dynamics. That is, $\gamma_{t+dt} = e^{-i\Ham dt}\gamma_t e^{i\Ham dt}.$ Note that, to second order, we have
\begin{align}\label{eq:gamma_approx}
\gamma_{t+dt}\approx{}& \left(1-i\Ham dt-\frac{1}{2}\Ham^2 dt^2\right) \gamma_t \left(1+i\Ham dt-\frac{1}{2}\Ham^2 dt^2\right)\notag\\
\approx{}& \gamma_t -i\Ham\gamma_t dt -\frac{1}{2}\Ham^2\gamma_t dt^2 \notag\\
&{} +i\gamma_t\Ham dt - \frac{1}{2}\gamma_t\Ham^2 dt^2 +\Ham\gamma_t\Ham dt^2\notag\\
={}&\gamma_t -i[\Ham,\gamma_t]dt - \frac{1}{2}(\Ham^2\gamma_t+\gamma_t\Ham^2-2\Ham\gamma_t\Ham)dt^2\notag\\
={}&\gamma_t -i[\Ham,\gamma_t]dt - \frac{1}{2}[\Ham,[\Ham,\gamma_t]]dt^2\notag\\
\equiv{}&\gamma_t - adt - \frac{1}{2}bdt^2.
\end{align}
Since the dynamics is unitary and $\gamma_t$ is full-rank, we have \(\ln\gamma_{t+dt} = \ln(e^{-i\Ham dt}\gamma_t e^{i\Ham dt}) = e^{-i\Ham dt} (\ln\gamma_t) e^{i\Ham dt}\), and the negative entropy is conserved as
\(
\Tr(\gamma_{t+dt}\ln\gamma_{t+dt}) = \Tr(e^{-i\Ham dt}\gamma_t e^{i\Ham dt}e^{-i\Ham dt} (\ln\gamma_t) e^{i\Ham dt}) =  \Tr(\gamma_{t}\ln\gamma_{t}).
\)
Therefore, $S(\gamma_{t+dt}\rVert\gamma_t)$ can be rewritten as
\begin{align*}
S(\gamma_{t+dt}\rVert\gamma_t) & = \Tr(\gamma_{t+dt}\ln\gamma_{t+dt}) - \Tr(\gamma_{t+dt}\ln\gamma_{t})\\
&= \Tr((\gamma_t-\gamma_{t+dt})\ln\gamma_t).
\end{align*}
By (\ref{eq:gamma_approx}), we can approximate $S(\gamma_{t+dt}\rVert \gamma_t)$ as
\[
S(\gamma_{t+dt}\rVert \gamma_t)\approx\Tr(a\ln\gamma_t)dt + \frac{1}{2}\Tr(b\ln\gamma_t)dt^2.
\] 
Note that 
\(
\Tr(a\ln\gamma_t) = i\Tr(\Ham\gamma_t\ln\gamma_t - \gamma_t\Ham\ln\gamma_t) = 0.
\) Hence we finally have \eqref{eq:S_in_speedlimit} as desired:
\begin{align*}\label{eq:S_dt2}
	S(\gamma_{t+dt}\rVert \gamma_t) \approx \frac{1}{2}\Tr([\Ham,[\Ham,\gamma_t]]\ln\gamma_t)dt^2.
\end{align*}
We need to verify that $S(\gamma_{t+dt}\rVert \gamma_t)\ge 0$ to second order in $dt$. Let the eigendecomposition of $\gamma_t$ be $\gamma_t=\sum_{i=1}\lambda_i\ket{i}\bra{i}$. Note the facts that $\Tr(\OP_A\OP_B)=\Tr(\OP_B\OP_A)$ and $\Tr(\OP_A[\OP_B,\OP_C])=\Tr([\OP_A,\OP_B]\OP_C)$ for generic operators $\OP_A,\OP_B$, and $\OP_C$. Therefore,
\begin{align*}
\Tr([\Ham,[\Ham,\gamma_t]]\ln\gamma_t)& =\Tr(\ln\gamma_t[\Ham,[\Ham,\gamma_t]])\\
&=\Tr([\ln\gamma_t, \Ham][\Ham,\gamma_t]).
\end{align*}
The $(i,k)$-th entry of $[\ln\gamma_t, \Ham]$ is 
\[
[\ln\gamma_t, \Ham]_{ik}= \bra{i} [\ln\gamma_t,\Ham]\ket{k}=(\ln\lambda_i-\ln\lambda_k)\bra{i}\Ham\ket{k}.
\] 
Similarly, the $(k,j)$-th entry of $[\Ham,\gamma_t]$ is
\[
[\Ham,\gamma_t]_{kj}=\bra{k} [\Ham,\gamma_t]\ket{j}=(\lambda_j-\lambda_k)\bra{k}H\ket{j}.
\] 
Therefore, we have
\begin{align*}
&\Tr([\ln\gamma_t, \Ham][\Ham,\gamma_t]) \\
={}& \sum_{i,k} (\ln\lambda_i-\ln\lambda_k)(\lambda_i-\lambda_k)\bra{i}\Ham\ket{k}\bra{k}\Ham\ket{i}\\
={}& \sum_{i,k} (\ln\lambda_i-\ln\lambda_k)(\lambda_i-\lambda_k)|\Ham_{ik}|^2,
\end{align*}
where $\Ham_{ik}$ is the $(i,k)$-th entry of $\Ham$. Since $(\ln x-\ln y)(x-y)\geq 0$ for $x,y>0$, we have \[\Tr([\ln\gamma_t, \Ham][\Ham,\gamma_t])\ge 0.\] This result verifies that $S(\gamma_{t+dt}\rVert \gamma_t)\geq 0$ to the order of $dt^2$.

\twocolumngrid


\begin{references}
\bibitem{Note}
In fact, for the inequality to hold for all $s>0$, it suffices that, in evaluating the trace term, the operator $\OP_c$ under $\gamma_0$ can be equivalently viewed as a classical random variable $O_c$, which is not necessarily bounded, under some properly defined probability measure $P_0$. Consequently, $\Tr(e^{\xi s \OP_c}\gamma_0)=\E_{O_c\sim P_0}[e^{\xi sO_c}]$, where the moment-generating function on the right-hand side is finite for all $s>0$ and $\xi=\pm 1$. The construction of $O_c$ and $P_0$ is given in Sec.~\ref{subsec:subGQFRI} and in Appendix~\ref{Appendix:Derivation_QFRI} below Eq.~\eqref{eq:GQFR}.
\end{references}
\end{document}